\documentclass[12pt]{article}
\usepackage[utf8]{inputenc}
\usepackage{amsmath,amssymb}
\usepackage{booktabs}
\usepackage{algorithm}
\usepackage{algorithmic}
\usepackage{hyperref}
\usepackage{geometry}
\usepackage{hyperref}
\usepackage{amsthm}
\usepackage{float}
\hypersetup{breaklinks=true}

\title{Exact Computation of the Catalan Number \(C(2{,}050{,}572{,}903)\)}
\author{Mahesh Ramani\\
\textit{Independent}}

\date{November 2025}

\begin{document}
\maketitle

\begin{abstract}
This paper presents a two-phase algorithm for computing exact Catalan numbers at an unprecedented scale. The method is demonstrated by computing \(C(n)\)
for \(n=2,050,572,903\) yielding a result with a targeted \(1,234,567,890\) decimal digits. To circumvent the memory limitations associated with evaluating large factorials, the algorithm operates exclusively in the prime-exponent domain. Phase 1 employs a parallel segmented sieve to enumerate primes up to \(2n\) and applies Legendre’s formula to determine the precise prime factorization of \(C(n)\).   The primes are grouped by exponent and serialized to disk. Phase 2 reconstructs the final integer using a memory-efficient balanced product tree with chunking. The algorithm runs on a time complexity of $\Theta(n (\log n)^2)$ bit-operations and a space complexity of $\Theta(n \log n)$ bits. This result represents the largest exact Catalan number computed to date. Performance statistics for a single-machine execution are reported, and verification strategies—including modular checks and SHA-256 hash validation—are discussed. The source code and factorization data provided to ensure reproducibility.
\end{abstract}
\section{Introduction}
Catalan numbers appear in many combinatorial problems (e.g.\ binary trees, polygon triangulations) 
\cite{stanley} and grow exponentially with \(n\), given their asymptotic growth $C(n)\sim \frac{4^n}{n^{3/2}\sqrt{\pi}}$.  By definition, 
\[
C(n)=\frac{(2n)!}{n!\,(n+1)!},
\]
cannot be computed directly for large \(n\) due to the astronomical size of \((2n)!\).  Traditional recursive or dynamic programming methods become infeasible for \(n\) on the order of millions, let alone billions.  Instead, a prime-factorization approach could be used: by Legendre’s formula \cite{legendre}, one can compute the exponent of each prime \(p\le2n\) in the numerator and denominator, thereby determining the exact prime-power factorization of \(C(n)\).  Grouping primes by exponent and reconstructing via balanced product trees yields the final integer.

This approach is implemented in two phases.  Phase~1 sieves primes up to \(2n\) using a parallel segmented sieve and computes each prime’s exponent in the factorization of \(C(n)\).  The primes are grouped by exponent and written to a factorization file.  In Phase~2, the file is read and processed group-by-group: for each exponent \(e\),
\[
\displaystyle Q_e=\prod_{p\in\text{group}_e} p
\]
 and \(P_e=Q_e^e\) are computed.  The final Catalan number is \(\prod_e P_e\), where the inner and outer products are computed using balanced product trees. To determine the specific input $n$ required to produce a Catalan number with exactly 1,234,567,890 decimal digits, the asymptotic approximation derived from Stirling's formula \cite{stirling}, 
\[
\displaystyle 
C(n) \sim 4^n / (n^{3/2}\sqrt{\pi}). 
\]
was inverted. By solving the transcendental equation, 
\[
\displaystyle
\log_{10} C(n) \approx n \log_{10} 4 - \frac{3}{2}\log_{10} n - \frac{1}{2}\log_{10} \pi \approx 1,234,567,889
\] the value $n = 2,050,572,903$ was identified as a valid candidate (after verification through computing the approximate value of \(\log C(n)+1\)). This pre-computation allowed precise targeting of the result's magnitude to ensure an interesting result.

This approach enabled the exact evaluation of \(C(2,050,572,903)\), a scale of computation shown to exceed the capabilities of general-purpose software like SageMath \cite{sagemath} and PARI/GP \cite{parigp}. Reproducibility is guaranteed through the provision of factorization files and cryptographic hashes. This achievement not only establishes a new benchmark for the largest known exact Catalan number but also validates a modular architecture applicable to any factorial-based sequence. Furthermore, the analysis highlights critical bottlenecks in integer reconstruction, offering specific benchmarks for the improvement of arbitrary-precision arithmetic libraries.

This paper presents the mathematical foundations (Section~\ref{sec:math}), algorithmic details (Section~\ref{sec:algorithm}), and implementation (Section~\ref{sec:impl}), along with pseudocode for each phase.  In Section~\ref{sec:results}  experimental results and resource usage are reported for the computation for \(n=2,050,572,903\). Section~\ref{sec:discussion} discusses bottlenecks, future improvements, and time/space complexities.  This paper is concluded by describing verification strategies, including the provided SHA-256 hash of the final result, to enable independent validation.
All code, data, and output files for this computation are available on GitHub \cite{ramani}.

\section{Mathematical Foundation}\label{sec:math}
Let \(v_p(m)\) denote the exponent of prime \(p\) in the prime-factorization of integer \(m\) (the \(p\)-adic valuation).  Legendre’s formula \cite{legendre} states that for positive integers \(m\) and prime \(p\),
\[
v_p(m!) \;=\; \sum_{k=1}^\infty \left\lfloor\frac{m}{p^k}\right\rfloor,
\]
a finite sum since \(\lfloor m/p^k\rfloor=0\) for \(p^k>m\). The \(p\)-adic valuation of the Catalan number can be expressed as:
\[
v_p(C(n)) \;=\; v_p((2n)!) \;-\; 2\,v_p(n!) \;-\; v_p(n+1).
\]
Thus each prime \(p\le 2n\) contributes an exponent \(e_p = v_p(C(n)),\)
and \(C(n)=\prod_{p\le2n}p^{\,e_p}\).  Note \(v_p(n+1)\) is nonzero only for primes dividing \(n+1\).  Because \(C(n)\) is known to be an integer (e.g.\ by combinatorial interpretation or Kummer’s theorem \cite{kummer}), all \(e_p\) computed as above are nonnegative.  Working with these exponents avoids forming the enormous factorials \((2n)!\), \(n!\), etc.

\section{Algorithm Overview}\label{sec:algorithm}

\subsection{Phase 1: Factorization Generation (\texttt{catalan.py})}
The set of all primes \(p\le 2n\) is generated using a parallel segmented Sieve of Eratosthenes  \cite{crandall}.  Implemented in Python, the sieve first identifies primes up to \(\le \sqrt{2n}\) to serve as seeds; subsequently, disjoint intervals are processed in parallel to maximize throughput on multi-core architectures. For each identified prime \(p\) found, the \(p\)-adic valuations
\[
a_p = v_p((2n)!),\quad b_p = v_p(n!),\quad c_p = v_p(n+1),
\]
are calculated using Legendre’s formula.  Then \(e_p = a_p - 2b_p - c_p\).  For primes \(n<p\le 2n\), the term \(v_p((2n)!) - 2v_p(n!)\) simplifies to either \(1\) or \(0\) (since \(\lfloor2n/p\rfloor - 2\lfloor n/p\rfloor\) is 1 if and only if \(p\) appears once in the binomial \(\binom{2n}{n}\)); the code handles this case automatically.  The script factors \(n+1\) by trial division using the prime list to attain \(c_p\).

Primes are subsequently grouped by their computed exponent \(e_p\).  Let \(G_e\) denote the set of all primes with exponent \(e\).  In the presented implementation,  the resulting factorization data is serialized to a text file (as \texttt{catalan\_[n]\_factorization.txt}; [n] represents the value of n) using the following structure:
\begin{verbatim}
# Prime factorization of Catalan(n)
# exponent=E count=K
p1 p2 p3 ...   (primes with exponent E)
...
\end{verbatim}
Each block lists all primes with the same exponent \(E\).  In parallel, the code accumulates a high-precision approximation of \(\log_{10}C(n)\) to estimate the digit count (which is then output as metadata). The pseudocode summarizing this phase is found in Algorithm~\ref{alg:phase1}.

\begin{algorithm}[H]
\caption{Factorization Generation for Catalan($n$) (in \texttt{catalan.py})}
\label{alg:phase1}
\begin{algorithmic}[1]
\REQUIRE Integer \(n\).
\STATE Compute all primes \( \le 2n\) via parallel segmented sieve.
\STATE Factor \(m = n+1\) by trial division with primes \(\le \sqrt{n+1}\).
\FOR{each prime \(p \le 2n\)}
  \STATE Compute \(a = \sum_{k\ge1}\lfloor 2n / p^k \rfloor\).
  \STATE Compute \(b = \sum_{k\ge1}\lfloor n / p^k \rfloor\).
  \STATE Set \(e = a - 2b\).
  \IF{\(p\) divides \(n+1\) with exponent \(c\)}
    \STATE Set \(e = e - c\).
  \ENDIF
  \IF{\(e > 0\)}
    \STATE Add \(p\) to group \(G_e\).
  \ENDIF
\ENDFOR
\STATE Write each group \(G_e\) to \texttt{catalan\_\(n\)\_factorization.txt}.
\STATE Estimate and output digit count of \(C(n)\).
\end{algorithmic}
\end{algorithm}

\subsection{Phase 2: Reconstruction (\texttt{catalanreconstructor.py})}
In the reconstruction phase, the pre-computed factorization data is ingested to assemble \(C(n)\).  The file is processed iteratively by exponent group. For each exponent \(e\) and corresponding prime set \(G_e\), the fundamental product
\[
Q_e = \prod_{p \in G_e} p.
\]
is evaluated. This operation employs a memory-efficient balanced product tree algorithm, wherein operands are recursively paired and multiplied to minimize the bit-length of intermediate values.  Following this, the term \(P_e = Q_e^e\) is computed utilizing optimized arbitrary-precision exponentiation routines (\texttt{gmpy2} \cite{gmpy2}).  Finally, the Catalan number is \(\prod_e P_e\), once again aggregating the partial results via a balanced product tree. The final integer is serialized to a binary file, with an optional decimal text output generated for verification. Binary serialization is utilized to avoid the significant overhead of decimal conversion and text I/O during intermediate steps.

Algorithm ~\ref{alg:phase2}  outlines this procedure, where \(\mathrm{BalancedProductTree}(S)\) denotes a recursive divide-and-conquer multiplication of elements in set \(S\).  Given the magnitude of the sets \(G_e\)  and the intermediate list  \(\mathcal{P}\), operations are batched into chunks (typically \ \(10^5\) elements at a time). A hybrid strategy employing periodic memory reclamation and disk-based offloading is implemented to strictly constrain RAM usage during the combination of these partial products.

\begin{algorithm}[H]
\caption{Integer Reconstruction for Catalan($n$) (in \texttt{catalanreconstructor.py})}
\label{alg:phase2}
\begin{algorithmic}[1]
\REQUIRE Factorization groups \(G_e\) from file.
\STATE Initialize list \(\mathcal{P} = []\).
\FOR{each exponent \(e\) with prime list \(G_e\)}
  \STATE Compute \(Q_e = \mathrm{BalancedProductTree}(G_e)\).
  \STATE Compute \(P_e = Q_e^e\) using fast pow.
  \STATE Append \(P_e\) to \(\mathcal{P}\).
\ENDFOR
\STATE Compute \(C(n) = \mathrm{BalancedProductTree}(\mathcal{P})\).
\STATE Write \(C(n)\) to \texttt{catalan\_\(n\)\_reconstructed.bin}.
\end{algorithmic}
\end{algorithm}

\section{Implementation and Reproducibility}\label{sec:impl}
The code is implemented in Python~3 using standard libraries.  Key files are:
\begin{itemize}
    \item \texttt{catalan.py}: Phase~1 factorization generation.
    \item \texttt{catalanreconstructor.py}: Phase~2 integer reconstruction.
\end{itemize}
The latter uses the \texttt{gmpy2} library (which wraps GMP \cite{gmp}) for multiprecision arithmetic.  The former uses \texttt{multiprocessing} for parallel sieve segments.

To reproduce the result, one should run \texttt{catalan.py} with \(n=2,050,572,903\), which will produce \texttt{catalan\_2050572903\_factorization.txt}.  Then run \texttt{catalanreconstructor.py} on that factorization file to produce \texttt{catalan\_2050572903\_reconstructed.bin}.  These output files are provided for reference.  All scripts include parameters (segment size, chunk size) with default values that worked for this run; these can be tuned for other hardware. More information on my computer specifications is found in Section~\ref{sec:results}.

\section{Results}\label{sec:results}
The two phases were executed sequentially on the specified hardware. Table~\ref{tab:stats} summarizes the key performance metrics.  Notably, the input \(n=2,050,572,903\) yielded a Catalan number with exactly \(1,234,567,890\) decimal digits (as targeted).  The final binary result file has size 512,643,220 bytes (= 512.643 MB decimal = 488.895 MiB).
The total wall-clock time was 1549.14~seconds, with 179.37~s in Phase~1 and 1369.77~s in Phase~2 (measured with \texttt{time}).  Peak memory usage reached approximately 4 GB during Phase 2; Phase 1 memory consumption was negligible in comparison.

\begin{table}[ht]
\centering
\caption{Computation statistics for $n=2,050,572,903$.}
\label{tab:stats}
{\footnotesize
\begin{tabular}{@{}ll@{}}
\toprule
Statistic & Value \\
\midrule
$n$ & $2,050,572,903$ \\
Decimal digits of $C(n)$ & $1,234,567,890$ \\
Bit length of $C(n)$ & $4,101,145,759$ \\
Size of result file & 488.9 MiB \\
SHA-256 of output & \texttt{dac68f4ee35db8e9400e68bd6140e6cbccec6fb8ce81059318400e2c44e45ae4} \\
Phase~1 runtime (wall) & 179.37~s \\
Phase~2 runtime (wall) & 1369.77~s \\
Total wall time & 1549.14~s \\
Peak memory (Phase~2) & 4~GB \\
Hardware & AMD Ryzen 9 4900HS, 16~GB RAM, 6~GB GPU (unused) \\
OS & Windows \\
Software & Python 3.11.9, gmpy2 \\
\bottomrule
\end{tabular}
}
\end{table}

These results demonstrate the feasibility of computing Catalan numbers in the billion-digit range with commodity hardware, especially considering the overhead that is inherently present with the use of interpreted languages in regards to number evaluation.  To measure the presented algorithm's efficiency (both memory and time wise) against open-source evaluation tools, wall times of different Catalan number evaluations were measured in Table ~\ref{tab:times}. This data was generated using the \texttt{timeit} function on Sage, which measures wall time over multiple trials, and averaging wall time reports on my local machine five times.
\begin{table}[ht]
\centering
\caption{Wall times for values of  \(C(n)\).}
\label{tab:times}
{\footnotesize
\begin{tabular}{@{}ll@{}ll@{}}
\toprule
Value of \(n\) & Proposed Algorithm (s) \indent & SageMath \cite{sagemath} ran Locally (s) \\
\midrule
$n=10^6$ & $2.923$ s & $0.378$ s \\
$n=10^7$ & $6.817$ s & $7.30$ s\\
$n=10^8$ & $62.835$ s & $134.0$ s \\

\bottomrule
\end{tabular}
}
\end{table}

Comparative analysis was also performed using PARI/GP (v2.15) \cite{parigp}. While PARI/GP exhibited superior performance for smaller values of \(n\), it encountered a memory exhaustion error at \(n \approx 1.34375 \cdot 10^9\). The standard recursive or factorial-based implementations in general-purpose CAS require loading intermediate values into RAM that exceed the 16 GB capacity of the test environment, highlighting the necessity of the disk-based, prime-exponent approach for inputs of this magnitude.

\section{Discussion} \label{sec:discussion}

The two-phase pipeline---factorization followed by reconstruction---is central to the algorithm's feasibility at this scale. Phase 1, the factorization, can be executed on machines with modest RAM, as its primary tasks are sieving for primes and computing Legendre's formula exponents. The dominant bottleneck, both in terms of computational time and memory, is the reconstruction phase. The large-integer multiplications required to combine the prime powers consume the majority of the runtime and dictate the hardware requirements for extreme-scale computations.

Disk I/O can also become a limiting factor during reconstruction, particularly when intermediate products are offloaded to disk to manage memory. For this reason, high-throughput storage such as NVMe or other solid-state drives is strongly preferred over traditional spinning disks to mitigate this potential bottleneck. Ultimately, the performance of Phase 2 is bound by the available RAM and the efficiency of the underlying multiprecision arithmetic library.

\subsection{Time Complexity}
\begin{itemize}
    \item \textbf{Sieve:} Enumerating primes up to $2n$ costs approximately $\mathcal{O}(n \log \log n)$ work. The sieve benefits significantly from parallelization.
    \item \textbf{Exponent Computation:} For each prime $p$, the computation of its exponent via Legendre's formula involves a sum of floor divisions. The cost of this step is subdominant compared to the reconstruction phase for the large values of $n$ targeted by this work.
    \item \textbf{Reconstruction:} This phase is dominated by large-integer multiplications. Let $B$ be the bit-length of the final integer, where $B = \Theta(n \log n)$. If $M(B)$ denotes the cost of multiplying two $B$-bit integers, the reconstruction cost is approximately $\mathcal{O}(M(B) \log P)$, where $P$ is the number of distinct prime factors. In practice, using GMP's \cite{gmp} advanced multiplication algorithms, this behaves asymptotically like $\Theta(n (\log n)^2)$ bit-operations.
\end{itemize}

\subsection{Space Complexity}
The peak memory usage is dictated by the largest intermediate integer held in memory during the balanced product-tree reductions in Phase 2. The size of this integer, and thus the space complexity, is determined by the bit-length of the final Catalan number, which is $\Theta(n \log n)$ bits. The storage for the prime numbers generated during the sieve requires $\mathcal{O}(n / \log n)$ machine words. The two-phase design effectively trades increased disk I/O for a reduction in peak RAM requirements by storing the complete factorization on disk before reconstruction begins.

\subsection{Generalizability}

While this work focuses on the Catalan sequence, the presented architecture is modular. Specifically, the reconstruction utility (\texttt{catalanreconstructor.py}) functions as a generic engine for computing products of prime powers, independent of the combinatorial source of the exponents. Consequently, the method extends to any integer sequence defined by ratios of products of factorials, including binomial coefficients $\binom{n}{k}$, multinomial coefficients, and super-factorials.

Formally, let $\mathcal{F}$ denote the class of combinatorial quantities expressible as:
\[
    \mathcal{F} = \frac{\prod_{i=1}^{k} (a_i!)}{\prod_{j=1}^{m} (b_j!)}
\]
where $a_i, b_j \in \mathbb{Z}^+$. The prime factorization of any $N \in \mathcal{F}$ can be determined by extending Legendre's formula. By the linearity of the $p$-adic valuation $v_p$, the exponent of any prime $p$ in the factorization of $N$ is:
\[
    v_p(N) = \sum_{i=1}^{k} v_p(a_i!) - \sum_{j=1}^{m} v_p(b_j!)
\]
Substituting Legendre's formula $v_p(n!) = \sum_{r=1}^{\infty} \lfloor \frac{n}{p^r} \rfloor$, we obtain:
\[
    v_p(N) = \sum_{i=1}^{k} \sum_{r=1}^{\infty} \left\lfloor \frac{a_i}{p^r} \right\rfloor - \sum_{j=1}^{m} \sum_{r=1}^{\infty} \left\lfloor \frac{b_j}{p^r} \right\rfloor
\]
The value $N$ is then reconstructed as the product over all relevant primes:
\[
    N = \prod_{p \le \max\left(\max_i a_i, \max_j b_j\right)} p^{v_p(N)}
\]
Phase 1 of the algorithm can be adapted to any member of $\mathcal{F}$ by simply modifying the summation logic for the exponents. Note that by integrality, the computed exponents \(v_p(N)\) are nonnegative; otherwise, negative exponents indicate that the expression is not integral.

Phase 2 remains unchanged. Assuming the use of Fast Fourier Transform based multiplication (e.g., the algorithm used by GMP for large operands), the reconstruction time complexity relies on the bit-length of the final integer, discussed in Section 6.1.
\section{Conclusion}

This paper presents a reproducible, memory-efficient framework for computing extreme-scale combinatorial numbers, demonstrated by the exact evaluation of $C(2,050,572,903)$. By operating exclusively in the prime-exponent domain, the algorithm circumvents the memory constraints that limit standard recursive implementations in general-purpose computer algebra systems.

As discussed in Section 6.3, this architecture is not limited to the Catalan sequence; the decoupled reconstruction engine generalizes to any integer defined by factorial ratios. This offers a viable pathway for high-precision investigations in broader areas of enumerative combinatorics using commodity hardware. The provided source code and factorization data enable independent verification and serve as a benchmark for optimizing large-integer arithmetic libraries.

Finally, Catalan numbers have many combinatorial interpretations; one classical interpretation yields a concrete result:

\begin{quote}
\textbf{Corollary.} \textit{The number of distinct ways to triangulate a convex polygon with 2,050,572,905 sides is exactly the integer provided in the title of this paper \cite{stanley}.}
\end{quote}

\end{document}